\documentstyle[12pt]{article}
\def\strut{\rule[-.5cm]{0cm}{1cm}}
\def\beq{\begin{equation}}
\def\eeq{\end{equation}}

\def\dspace{\baselineskip =.30in}

\begin{document}
\begin{titlepage}
\begin{center}
\rightline{UNIL--TP--5/95}
\rightline{BA--95--17}
\rightline{hep-ph/9512345}
\vskip1.5truecm
{\Large \bf Radiative Electroweak Breaking with Pseudogoldstone
Higgs Doublets}\\
\vskip.75truecm
{{\large \bf B. Ananthanarayan\footnote{Present address:  Institut f\"ur
Theoretische Physik, Universit\"at Bern, CH 3012, Bern, Switzerland}}\\
Institut de Physique Th\'eorique,
Universit\'e de Lausanne, \\
CH-1015 Lausanne, Switzerland\\
\vskip.5truecm
{\large \bf Q. Shafi}\\
Bartol Research Institute, University of Delaware\\ Newark, DE 19716, USA\\}

\end{center}
\vskip.75truecm
\bigskip
\bigskip

\begin{abstract} We consider a realistic example of supersymmetric grand
unification based on $SU(3)_c \times SU(3)_L \times SU(3)_R$ in which the
electroweak (EW) higgs doublets are `light' as a consequence of the
`pseudogoldstone' mechanism.  We discuss radiative EW breaking in this model,
exploring in particular the `small' (order unity) and `large' $(\approx
m_t/m_b)$ $\tan \beta$ regions by studying the variations of $r (\equiv
\sqrt{\mu^2_{1,2}/\mu^2_3})$, where $\mu^2_{1,2,3}$ are the well known MSSM
parameters evaluated at the GUT scale.  For $r$ sufficiently close to unity
the quantity $\tan \beta$ can be of order unity, but the converse is not
always true.

 \end{abstract} \end{titlepage}

\dspace
\section{Introduction}

Understanding how the electroweak higgs doublets of the minimal supersymmetric
standard model (MSSM) remain `light' $(\sim 10^2 GeV)$ within the framework of
supersymmetric grand unified theories (SUSY GUTS) poses an important challenge
for model builders.  In supersymmetric trinification with gauge group $G
\equiv SU(3)_c \times SU(3)_L \times SU(3)_R$, by imposing suitable discrete
symmetries for instance, it is possible to protect the EW doublets from
becoming superheavy without fine tuning [1].  The supersymmetric $\mu$-term of
MSSM arises from a higher order (non-renormalizable) term in the
superpotential.  This approach leads to a number of testable predictions.  The
proton turns out to be essentially stable, while the MSSM parameter $\tan
\beta \approx m_t/m_b$.  It is interesting to recall that in this case, by
fixing $m_b (m_b) = 4.25 \pm 0.10$ GeV and $\alpha_s (M_z) = 0.12 \pm 0.01$,
the top quark mass was predicted [2] to lie in a range which is in very good
agreement with the subsequent CDF/DO measurements.

A somewhat different approach for obtaining the light doublets relies on the
idea of an accidental `pseudo-symmetry' [3] which is spontaneously broken.  [It
also may be broken both explicitly as well as by radiative corrections.]
Examples [4,5]
based on SU(6) (SU(5) and SO(10) do not seem to work) and more recently [6] on
$G(\equiv (SU(3))^3)$ have been presented.  In this paper we wish to focus on
the pseudogoldstone mechanism in G and study the implications of merging it
with the radiative EW breaking scenario. In section 2  we provide the details
of this mechanism within the framework of G. What partially distinguishes this
example from some previous work based on $SU(6)$
 can be explained in terms
of the parameter $r \equiv \sqrt{\mu^2_1/\mu^2_3}\ (\equiv
\sqrt{\mu^2_2/\mu^2_3})$, where $\mu^2_1, \mu^2_2, \mu^2_3$  are the well
known mass squared parameters of the tree level scalar potential of MSSM,
evaluated at the GUT scale $M_G$. In the simplest $SU(6)$ model $r$ is equal
to unity, up to corrections of order $(1TeV/M_G)^2$, where 1 TeV specifies the
supersymmetry breaking scale.  In the $(SU(3))^3$ case, $r$ deviates from
unity even in the supersymmetric limit due to the presence of a superpotential
term which breaks pseudosymmetry at tree level.  Nonetheless, this leads to
the desired higgs doublets [6].  Indeed, in the absence of this additional
term $r$ is unity, but then the  top quark turns out to be massless at tree
level which is unacceptable. In section 3 we consider  radiative EW breaking
as well as the ensuing sparticle spectroscopy, focusing on $r$ very close to
unity such that $\tan \beta$ is of order unity.  We find interesting
constraints on the parameters, namely $\mid M_{1/2} \mid \
\stackrel{_<}{_\sim} \ m_0\ \stackrel{_<}{_\sim} \ \mid A \mid$, where
$M_{1/2} (m_0)$ denote the universal gaugino (scalar) mass, and $A$ is the
universal trilinear scalar coupling.  Figures 1-6 highlight this region of the
parameter space.  In Figs. 7-10 we show how by varying the
ratio $A/m_0$, the quantity $r \gg 1$ without $\tan \beta$ becoming large.  In
section 4 we briefly summarize the large $\tan \beta$ case obtained by varying
$r$ further away from unity (Fig. 11).

\section{The $(SU(3))^3$ Pseudogoldstone Model}

We consider a supersymmetric grand unified model
based on the gauge group $G\ \equiv \ SU(3)_c \ \times
SU(3)_L
\times SU(3)_R$. The matter (lepton, quark, antiquark) fields of the model
transform as
$(1, \bar{3}, 3), \ (3,3,1)$ and $(\bar{3}, 1, \bar{3})$ under G:

\newpage

\begin{eqnarray} \lambda_i & \ \equiv & \ \left( \begin{array}{ccc} H_1 &
H_2 & L \\ e^c &
\nu^c & N \end{array} \right)_i \nonumber \\ Q_i & \ \equiv & \ \left(
\begin{array}{c} u \\ d \\ g
\end{array} \right)_i \nonumber \\ Q^{(c)}_i & \ \equiv & \ (u^c\ d^c\
g^c)_i  , i\
= \ 1,2,3 \end{eqnarray}
The superfields $H_1, H_2, L$ are $SU(2)_L$ doublets,
where
$SU(3)_L\ (SU(3)_R)$ acts along the columns (rows) of the matrices in (1).
Under $SU(2)_L \times U(1)$, $H_{1i}, H_{2i}$ have the same quantum numbers as
the EW doublets, while $L_i$ denote the lepton doublets.

In order to break the gauge group G down to MSSM, we need higgs superfields
that transform as the $\lambda_i$'s in (1). The minimum number that is needed
is
two which we  denote as

\beq
\lambda (\bar{\lambda}) \
{\rm and} \ \lambda^{\prime} (\bar{\lambda}^{\prime}) \eeq
The conjugate superfields $\bar{\lambda}$ and $\bar{\lambda}^{\prime}$ are
needed
to preserve SUSY when G breaks to the standard model gauge group.
The scalar components of $\lambda(\bar{\lambda})$ acquire large non-zero vevs
along the
$N(\bar{N})$ directions such that G breaks to $SU(3)_c \times SU(2)_L
\times SU(2)_R
\times U(1)_{B-L}$. With $\lambda^{\prime } (\bar{\lambda}^{\prime })$
acquiring
large vevs
along the $\nu^{c \prime} (\bar{\nu}^{c \prime })$ direction, the resulting
unbroken symmetry  will be $SU(3)_c \times SU(2)_L \times U(1)$.

Let us begin by specifying the part of the
superpotential that involves the chiral
superfields $\lambda , \bar{\lambda}$:

\beq W_{\lambda} \ = \ S(\lambda \bar{\lambda}\ - \ \mu^2)\ + \ a
\lambda^3\ + \ b
\bar{\lambda}^3 \eeq
Here $\lambda$ stands for $\lambda^A_{\alpha},\ \lambda^3\ \equiv\
\epsilon_{ABC}\
\epsilon^{\alpha \beta \gamma } \lambda^A_{\alpha} \lambda^B_{\beta}
\lambda^C_{\gamma}(etc.)$, and S denotes a gauge singlet field S.
To see how the pseudogoldstone mechanism operates, consider a situation in
which we
include an analogous term $W_{\lambda '}$ for the $\lambda '$ sector, but
 there is no
$\lambda -
\lambda^{\prime}$ mixing [for details see Ref.[6]].
In this limit there appears a
larger global symmetry (``pseudo-symmetry")
\beq
G_{gl} \ = \ [SU(3)_c \times SU(3)_L \times SU(3)_R]_{\lambda} \ \otimes \
[SU(3)_c \times  SU(3)_L \times SU(3)_R]_{\lambda^{\prime}} \eeq
under which $W_\lambda +W_{\lambda '} +$ h.c. is invariant.
It has been shown [5] that when G breaks to $SU(3)_c \times SU(2)_L \times
U(1)$,
there  emerge a pair of `massless' doublets with the quantum numbers of the EW
higgs:
\begin{eqnarray}
P \ & = \ L \langle \nu^{c \prime} \rangle \ - \ H^{\prime}_2 \ \langle N
\rangle \nonumber \\
\bar{P} \ & = \ \bar{L} \langle \bar{\nu}^{c \prime} \rangle \ - \
\bar{H}^{\prime}_2 \ \langle
\bar{N} \rangle \end{eqnarray}
We observe that the $H^{\prime}_2$ component of $P$ has the correct quantum
numbers to couple (at tree level) to the down quarks and the
charged leptons. However, the corresponding component $\bar{H}^{\prime}_2$ of
$\bar{P}$
cannot serve as the second (`up' type) higgs doublet since it is forbidden from
 having a
renormalizable coupling to the quark superfields.  In particular,
the top quark is massless at tree level!

The resolution of this lies in
extending the field content of the model by including an
additional higgs
supermultiplet $\lambda^{\prime \prime} (\bar{\lambda}^{\prime
\prime})$. Consider the superpotential couplings $(M \sim M_{GUT})$ \beq
M \lambda^{\prime \prime} \bar{\lambda}^{\prime \prime} \ + \ f
\bar{\lambda}^{\prime}
\bar{\lambda} \bar{\lambda}^{\prime \prime} \eeq
The second term in (6) explicitly breaks $G_{gl}$ but in such a way that
the  desired `massless' pair survives.  A straightforward calculation
shows that the combination

\beq
-\sin \alpha\  H^{\prime \prime}_1 \ + \ \cos \alpha\ \bar{P} \eeq
is the required `up-type' higgs doublet. Here $\sin \alpha \ = \ z/(M^2 +
z^2)^{1/2}$ $(z = f(N^2 + \nu^{c^{\prime 2}})^{1/2})$ provides a
measure of the
breaking of the pseudosymmetry $G_{gl}$.

In order to evaluate the scalar potential involving the EW higgs doublets, we
turn attention to the relevant part of  the superpotential

\beq
W\ = \ \Sigma_i \ (a_i\ S_i)P\bar{P} + \stackrel{\sim}{W}\!(S_i, other
fields) \ + \ z
\bar{P} \bar{H}^{\prime \prime}_1 \ + M H^{\prime \prime}_1\
\bar{H}^{\prime \prime}_1
\eeq
where $S_i$ denote the $SU(2)\times U(1)$ singlet superfields. The scalar
$\rm{mass}^2$ matrix, after including the soft supersymmetry breaking
couplings,
is given  by $(m_0$ denotes the soft SUSY breaking scalar mass parameter and
$S_i$ in (9) and (10) denote the appropriate vev):

\beq
\begin{array}{ccccc}
& P^* & \bar{P} & \bar{H}^{\prime \prime *}_1 & H^{\prime \prime}_1
\nonumber\strut\\ P &
\mid a_i S_i \mid^2 + m^2_0 & a_i \frac{\partial
\stackrel{\sim}{W}^*}{\partial S_i^*}
+Aa_i S_i & a_iS_iz & 0 \nonumber\strut\\ \bar{P}^* & a^*_i
\frac{\partial
\stackrel{\sim}{W}}{\partial S_i} + A a^*_i S^*_i & \mid a_i S_i
\mid^2 + m^2_0
+ z^2 & Bz & Mz \nonumber\strut\\ \bar{H}^{\prime \prime}_1 & a_i
S_i z &
Bz & z^2+M^2 & B M \nonumber\strut\\ H^{\prime \prime *}_1 &
0 & Mz &
B M & M^2 \end{array}
\eeq
The presence of the `massless'
state (for $z = 0$) leads to the following relation

\begin{eqnarray} \mid a_i S_i \mid^2 \ + \ m^2_0 \ & = & \ a_i
\frac{\partial
\stackrel{\sim}{W}}{\partial S^*_i} \ + \ A a_i S_i \nonumber \\ &
= & m^2_0\
(b^2 + 1) \end{eqnarray}
with $b = \frac{C-A}{2m_0}$, where $A, B, C$ denote the
common tri-linear, bi-linear and linear scalar couplings from the soft SUSY
breaking at
$M_G$. Note that $\langle S_i \rangle = 0 (=bm_0)$ before (after)
SUSY breaking.

The $4 \times 4$ matrix in (9) can be simplified in a relatively
straightforward
manner and we will
focus on the `light higgs' sector which is given by the following $2 \times 2$
submatrix:

\beq
\begin{array}{cccc}
& P^* && H_u \nonumber \strut\\
P & m^2_0 (b^2 \cos^2 \alpha + 1) && m^2_0 (b^2 + 1) \cos \alpha
\nonumber
\strut\\
H^*_u & m^2_0 (b^2 +1) \cos \alpha && m^2_0 (b^2 \cos^2 \alpha + 1)
\end{array}
\eeq

where $H_u$ stands for the state given in eq.(7).

The following remarks are in order: \begin{description} \item[i.] With $\alpha
= 0$ the pseudosymmetry $G_{gl}$ is unbroken at tree level in the scalar
sector and we  have a pair of `massless' states with $\mu^2_1 \ =\ \mu^2_2 \
=\ \mu^2_3$ (at $M_G$). \item[ii.] The realistic case requires $\alpha \ \neq
\ 0$ so that, at $M_G$, \begin{eqnarray} \mu^2_1 \ = \ \mu^2_2 \ & = & \ m^2_0
(b^2 \ \cos^2 \alpha\ +\ 1) \nonumber \\ \mu^2_3 \ & = & \ m^2_0 (b^2 +1)\cos
\alpha \end{eqnarray} The deviation from unity (at $M_G$) of the ratio $r
\left( \equiv \sqrt{\mu^2_{1,2}/\mu^2_3} \right)$, which can be significant as
a consequence of (12), will be used in conjunction
with radiative electroweak breaking, to explore the parameter space of MSSM.
\item[iii.] In minimal supergravity, $B = A-m_0, C = A -2m_0$,  such that $b =
-1$. \end{description}

\section{Radiative Electroweak Breaking and
 $r \approx  1$}

In this section we wish to explore how close to unity $r$ can get without
running into conflict with the radiative electroweak breaking scenario.  For
$r$ sufficiently close to unity the well known parameter $\tan \beta$ turns
out to be of order unity.  The converse, however, is not necessarily true as
we will later see. The procedure we follow rests on minimizing the
renormalization group improved tree-level potential at a scale $Q_0\ \sim \
0.5 - 1 \ TeV$. The soft SUSY breaking parameters at this scale are estimated
through their one-loop evolution equations. The  reliability of minimizing the
tree-level potential in this manner has previously been studied [2] and yields
results that are consistent with minimizing the one-loop effective potential.

A knowledge of $\alpha_s$ and the electroweak couplings at present energies
enables us to estimate $M_G$ through their one loop evolution equations, with
supersymmetry breaking scale assumed to be of order $Q_0$. By specifying $h_t$
and $h_b (=h_{\tau})$  at $M_G$, one evolves the coupled system for the gauge
and third generation Yukawa  couplings down to $Q_0$, and solves for $\tan
\beta$ from the known value of $m_{\tau} \  (=1.78\ {\rm GeV})$. [We remark
that in the $(SU(3))^3$ framework the asymptotic relation
$h_b=h_\tau$ may be expected to hold near the Planck scale where the full $E_6$
theory is effective.] Furthermore, given $M_{1/2}, \ m_0$ and $A$ one obtains
the values  of $m^2_{H_2}$ and $m^2_{H_1}$ at $Q_0$, and from the minimization
conditions and  knowledge of $\tan \beta$, solves for $\mu (Q_0)$ and $B
(Q_0)$.  [Note that at one-loop  level, $\mu$ and $B$ do not enter the
evolution equations of the  remaining parameters.]  Knowing $\mu (Q_0)$ and $B
(Q_0)$, we can compute the  physical spectrum since the remaining parameters
are already known via their  evolution equations. If the result is a
consistent, stable $SU(2)\times U(1)$ breaking vacuum [one  that does not
conflict any phenomenological constraint]  we evolve $\mu (Q_0)$ and $B (Q_0)$
back to $M_G$ and evaluate the parameter $r$. The procedure we follow gives
results that are within $1-2\%$  of a one-loop calculation [7] which
explicitly reports the values of $\mu(M_X)$ and $B(M_X)$ in order to implement
a successful radiative breaking scenario.

We have performed a search in the parameter space spanned by \beq (h_t,\ h_b,\
M_{1/2},\ m_0,\ A) \eeq fixing $M_G\ \simeq\ 2 \times 10^{16}$ GeV, $Q_0\
\sim\ 350$ GeV, $\alpha_s (M_Z)\ =\ 0.12,\ \alpha_G\ = 1/25\ (\alpha_{em}
(M_Z)\ =\ 1/128)$. We illustrate the behaviour of the solutions in the desired
regions of the parameter space through several figures to bring out the
salient features. We begin by specifying a convention in which the Yukawa
couplings, $M_{1/2}$ and $\tan \beta$ are positive, allowing $A$ and $\mu$ to
be of either sign. We shall see that if $r$ is to be as close to unity as
possible, the parameters $A$ and $\mu$ will be required to have a common sign.
[The qualitative trends are similar when $\mu,\ A\ <\ 0$ is replaced by
$\mu,\ A\ >\ 0$.] In particular, the hierarchy which emerges, $M_{1/2}\ \ll\
m_0\ \ll\ \mid A \mid$, favours the sign of $A$ to be $+(-)$ when $\mu\ >\
0(<0)$.  The numerical choices for (13) correspond to those that yield
phenomenologically acceptable solutons often lying in the ranges reviewed in
Ref. [2].  Such solutions are typical and the gross features of the solutions
are perturbed in only a minor way when these are modified.

In Fig. 1 we illustrate the dependence of $r$ on $\tan \beta$, varying the
input value of $h_t (M_G)$. Notice that as $h_t$ increases it becomes harder
to achieve $r \approx 1$.

In Fig. 2 we illustrate the correlation between the input value of $A$
with the value of the parameter $r$. Here we have chosen $\mu\ <\ 0$ and one
finds that $A\ =\ -3m_0$ makes $r$ close to unity than say
 $A\ =\ -2m_0$, with all other parameters held fixed. [For $\mu\ >\
0$, it is $A\ =\ 3m_0$ versus say $A \ =\ 2m_0$.] Thus the requirement of
$r \ \approx \ 1$ favors a larger ratio for $\mid A \mid /m_0$.

In Fig. 3, we show the correlations between $M_{1/2}$ and $m_0$
when $r$ is plotted as a function of $\tan \beta$ for differing ratios
$m_0/M_{1/2}$.  The net conclusion to be drawn is that the $r\ \approx\ 1$
scenario enforces the correlation

\begin{eqnarray}
M_{1/2}\ \ll\ m_0\ & \ll & \ \mid A \mid \nonumber \\ {\rm and}
\hspace{.75in}{\rm sign}\ (A)\
& = & \ {\rm sign}\ (\mu) \end{eqnarray}

In Fig. 4 we further develop the message found in Fig. 1 for larger values of
$h_t$,
with $M_{1/2}/m_0$ and $A/m_0$ in the regimes singled out by the scenario, to
estimate how close to unity $r$ can get. We see
that to obtain $r\
\approx\ 1.05$ with $\mu\ <\ 0$ one requires $m_0$ to be as large as
$2M_{1/2}$.
Note that if $\tan \beta$ is too close to unity, the relation

\beq
m_t(m_t)\ =\ h_t(m_t)(174) \sin \beta
\eeq
may cause the top quark mass in the theory to come into conflict with the
CDF/D0
values [8].

With $\mu\ >\ 0$ a plot of $r$ as a function of $\tan
\beta$ is illustrated in Fig. 5.

The results above essentially  emerge due to the correlations enforced by the
well known evolution equations for the parameters  $\mu$ and $B$ [see Ref. 9]
and are given here for completeness:

 \begin{eqnarray} & \displaystyle
{{d\mu}\over{dt}}={\mu \over {16 \pi^2}}(-3g_2^2-\frac{3}{5}g_1^2
+h_\tau^2+3h_b^2+3h_t^2) & \nonumber\\ & \displaystyle {{dB}\over
{dt}}={1\over{8\pi^2}}(-3g_2^2M_2
-\frac{3}{5}g_1^2M_1 +h_\tau^2A_\tau + 3
h_b^2A_b+3h_t^2 A_t) & \end{eqnarray} where $t=\log Q/M_X$.

So far we have considered the variation of $r$ as a function of $\tan \beta$
in a region where $m_t(m_t)$ depends linearly on $\sin \beta$, namely where
$h_b(=h_{\tau})\ \ll\  h_t\ $ and is therefore neglected. Nevertheless, as
$\tan \beta$ increases, $h_b$ begins  to grow in relative importance and
eventually plays a role in arresting the growth of $m_t(m_t)$ as $\tan \beta$
grows for fixed $h_t$, eventually causing it to turn around.  This is the
reason why the quasi `infrared fixed point' prediction for $m_t$ (with $\tan
\beta \ \simeq\ m_t/m_b$) is significantly smaller than the corresponding
prediction with $\tan \beta\ \simeq\ 1$. For each value of $h_t$, with the
favoured hierarchy corresponding to $r \approx 1$, one can plot $r$ as a
function of $m_t(m_t)$. The result is presented in Fig. 6 describing the
correlation between $m_t(m_t)$ and $r$.  The cross-over for the two contours
$h_t\ =\ 2\ \rm{and}\ 1$ shows that,  provided the top is heavy enough, merely
lowering $h_t$ will not  suffice to enforce $r$ in the vicinity of unity.
Furthermore, from the preceding discussion, with $h_t\ =\ 1$ one cannot have a
top quark heavier than 182 GeV.

The conclusion to be desired from Fig. 6 is that should the top weigh more
than 180
GeV, $r\ \stackrel{<}{\sim}\ 1.12$ would be ruled out. If $m_t(m_t)\
\stackrel{>}{\sim}\ 191\
GeV$, we would be forced to have $h_t (M_G)\ \stackrel{>}{\sim}\ 2$ and $r\
\stackrel{>}{\sim}\ 1.24$. In this region the infrared prediction begins to
be realized,
whereby ever larger $h_t(M_G)$ would be implied with a rapidly increasing
lower bound
on $r$.
We also note here that with $\tan\beta \simeq 1.2$ and with the choice of
parameters of
Fig. 6, the $r\approx 1$ solution also satisfies the boundary condition of
the `minimal'
K\"ahler model, viz., $B=A-m_0$.

The main result to be drawn
from Figs. 1-6 is that with $r$ sufficiently close to
unity, the hierarchy $M_{1/2}\ \stackrel{_<}{_\sim} \ m_0\ \stackrel{_<}{\sim}
\ \mid A \mid$ is singled out.  In particular for $r$ sufficiently close to
unity $(\leq 1.15)$, one finds that
$1\ \stackrel{<}{\sim}\ \tan \beta\ \stackrel{<}{\sim}\ 5$.

It is reasonable to enquire if the `small' (order unity) $\tan \beta$ region
requires that $r$ also be close to unity.  This turns out to be not the case.
In Figs. 7a,b,c we show plots of $r$ versus $A/m_0$ for a typical choice of
$M_{1/2},\ m_0$ and $h_t$, with $\tan \beta$ varying between `order unity' to
`intermediate' values.  The parameter $\mu > 0$.  We see from 7b, for
instance, that $r$ can be large with $\tan \beta = 3$. This is a result of the
fact that the parameter $B(M_G)$ estimated through its one-loop evolution
equation suffers a change in sign as the ratio $A/m_0$ is varied from its
phenomenologically allowed lower bound of -3 for such values of $\tan\beta$.
For smaller values of $\tan\beta$, for instance 1.2, such a sign shift occurs
at values of this ratio smaller than --3, which are phenomenologically
excluded.  Furthermore, in order to demonstrate that these features are not a
result of accidental correlations between the input parameters, we present in
Figs. 8 and 9 systematic studies of the variations of $r$ as a function of
$A/m_0$ for differing choices of input parameters. These trends persist if
$\mu <0$ and $A/m_0 \rightarrow - A/m_0$.  An example is presented in Fig. 10.

\section{Large $\tan \beta$ versus $r$}

The phenomenological considerations are somewhat different in the event of
large $\tan \beta$ since the effects of $h_b$ and $h_\tau$ are no
longer negligible which tends to make the lighter stau approach the mass of
the LSP. Naturally we require this scalar tau to be heavier than the LSP.
Furthermore, in this limit, due to the essential degeneracy of $m_{H_1}^2$  and
$m_{H_2}^2$, $m_A$ also tends to remain low. In fact in the limit that $\tan
\beta\approx m_t/m_b$, such considerations play a crucial role in constraining
regions of the parameter space[11]. It is no longer possible to choose $m_0$
to be (much) larger than $M_{1/2}$, and the ratio $A/m_0$ is also forced to
remain rather low. We have performed a search in the parameter space
to minimize $r$ under these conditions. The result is displayed in Fig. 11.
Here we present the variation of $\tan \beta$ with $r$, obtained by varying
$h_b(=h_\tau)$ from $0.5 h_t$ to $h_t$, with $h_t$ chosen to be sufficiently
large (=$1.5$), such that $m_t(m_t)$ lies between $185$ and $181\ GeV$. The
universal gaugino mass $M_{1/2}$ is chosen to be $800\ GeV$ (and $Q_0\sim 1\
TeV$) in order to saturate the upper bound on the (bino-like) lightest
neutralino mass of $350\ GeV.$ For this figure we obtain the minimum value of
$r$ with $\mu>0$, with the maximum realizable values of $m_0$ and the ratio
$A/m_0$ consistent with the phenomenological requirements $m_A\geq m_Z$ and
$m_{\tilde{\tau}_1}\geq m_{\tilde{N}}$. What we find is that $r$ cannot be
smaller than about $1.5$ as we near the condition of exact Yukawa unification
$(h_t = h_b = h_{\tau}\ \rm{at}\ M_G)$.  Indeed if the exact Yukawa
unification condition is relaxed, there is considerable freedom in the ratio
$A/m_0$ as well, and even in the large $\tan\beta$ case $r$ can be just about
as large as one wants like in the `intermediate' $\tan\beta$ case.

\section{Conclusions}

The idea that the electroweak higgs doublets of MSSM may arise as
`pseudogoldstones' of an underlying supersymmetric grand unified theory can be
neatly realized within the framework of $SU(3)_c\times SU(3)_L\times
SU(3)_R$. In this work we have studied the implications when this idea is
merged with that of radiative electroweak breaking. In particular, we have
explored the constraints on the `universal' parameters $M_{1/2}, \ m_0$ and
$A$.  An important lesson is that the low energy parameter $\tan \beta$ can
vary all the way from order unity to $m_t/m_b$ in this class of models.
Depending on the top quark mass, certain lower bounds on the parameter $r$
have been identified.

\section*{Acknowledgements}

We thank Gia Dvali for
important discussions on the pseudogoldstone phenomenon in
SUSY GUTS.
B.A. thanks the Swiss National Science Foundation for support during the
course of this
work. The work of Q.S. is supported in part by the US Department of
Energy, Grant No.DE--FG02--91 ER 40626.

\section*{Note Added}
The results of this paper were briefly discussed at the
SUSY '95 meeting in Paris and at the``European High Energy Physics" conference
in Brussels.  After this paper was completed we came across a recent paper by
C. Cs\'{a}ki and L. Randall (hep-ph/9512278) in which similar ideas are
discussed.  Where our work overlaps the results are in broad agreement.

\newpage

\noindent{\bf Figure Captions}

\begin{enumerate}
\item Plot of $r\ \rm{versus}\ tan \beta$ for $h_t\ =\ 1$ and 0.8, with
$m_0 = 1.5 M_{1/2},\ A = -2m_0,\ \mu < 0$

\item Plot of $r$ vs. $\tan \beta$ for $A\ =\ -3
m_0$ and $-2
m_0$, with $m_0\ =\
1.5 M_{1/2}$ and $h_t \ =\ 1,\ \mu \ <\ 0$

\item Plot of $r$ vs. $\tan \beta$ for $m_0\ =\ 1.5 M_{1/2}$ and
$0.75 M_{1/2}$, with $A\ =\
-3 m_0$ and $h_t\ =\ 1, \ \mu\ <\ 0$

\item Plot of $r$ vs. $\tan \beta$ for $h_t\ =\ 2.5$ and $1.5, \
m_0\ =\
2\ M_{1/2}$ and $A\ =\
-3\ m_0,\ \mu\ <\ 0$

\item Plot of $r$ vs. $\tan \beta$ for $h_t \ =\ 1.5, \ M_{1/2}\ =\ 270
{\rm GeV},\ m_0\ =\ 340
{\rm GeV}$ for $A\ =\ -3
m_0,\ \mu\ <\ 0$ and $A\ =\ 3\ m_0,\ \mu\ <\ 0$

\item Plot of $r$ vs. $m_t(m_t)$ for $h_t\ =\ 1,2,3 ,\ M_{1/2}\ =\ 270
{\rm GeV},\ m_0\ =\ 340
{\rm GeV},\ A\ =\ 3m_0,\ \mu\ >\ 0$.

\item  (a) Plot of $r$ vs. $A/m_0$, for $M_{1/2}=280 {\rm GeV}$,
$m_0=340 {\rm GeV}$, $h_t=2.5$, $\tan\beta=1.2$, $\mu>0$, (b) As in (a)
with $\tan\beta=3.0$, and (c) as in (a) with $\tan\beta=7.8$.

\item As in Fig. 7 with $m_0=170{\rm GeV}$.

\item As in Fig. 7 with $M_{1/2}=420 {\rm GeV}$ and $m_0=510{\rm GeV}$.

\item As in Fig. 9c with $\mu<0$.

\item Plot of $r$ vs. $\tan \beta$ in the large $\tan \beta$
 regime with $m_0$ and
$A/m_0$ chosen optimally so as to minimize $r$ and saturate
 the requirement that
$m_A\geq m_Z$ and $m_{\tilde{\tau}_1}\geq m_{\tilde{N}}$.

\end{enumerate}

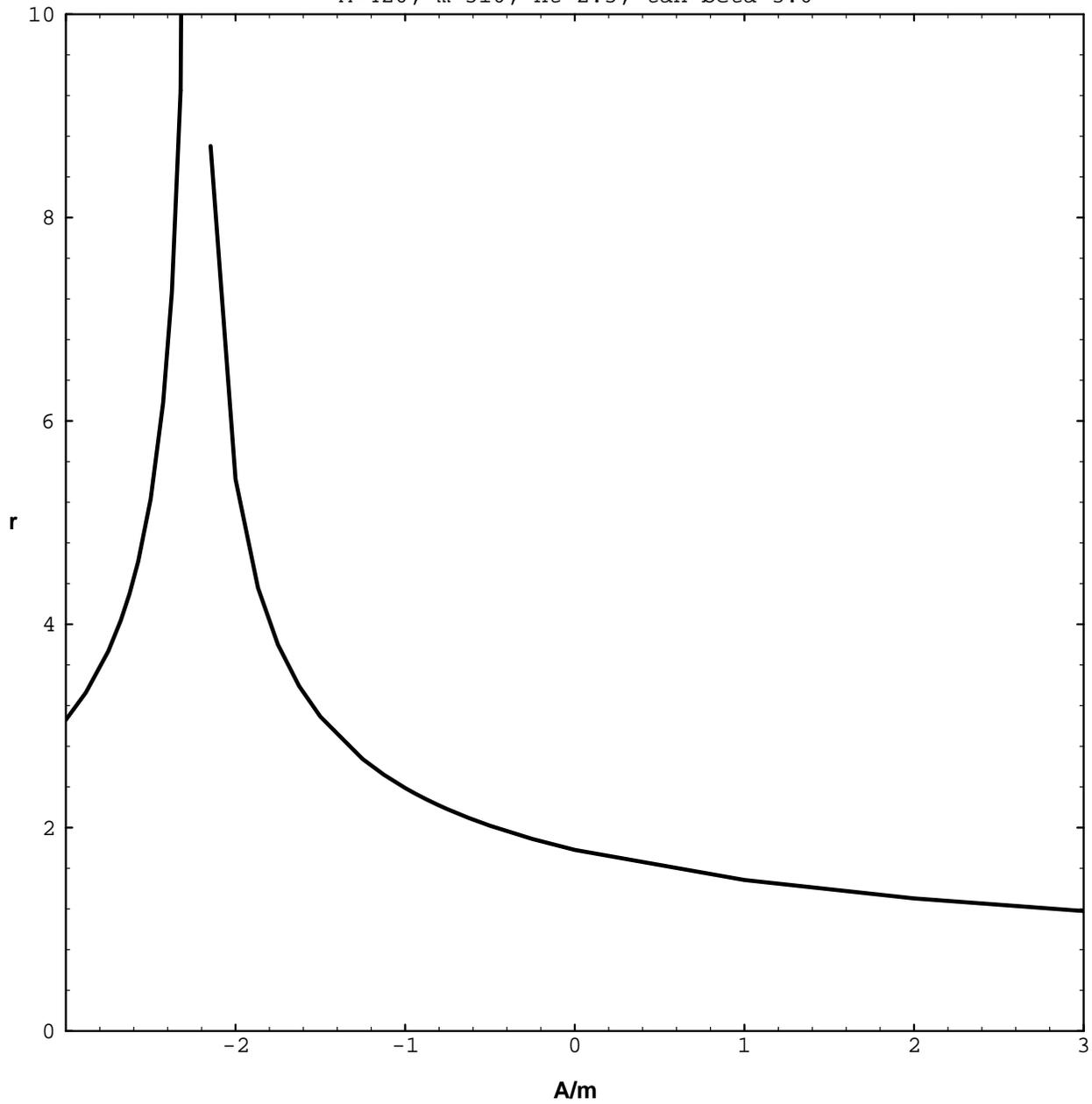
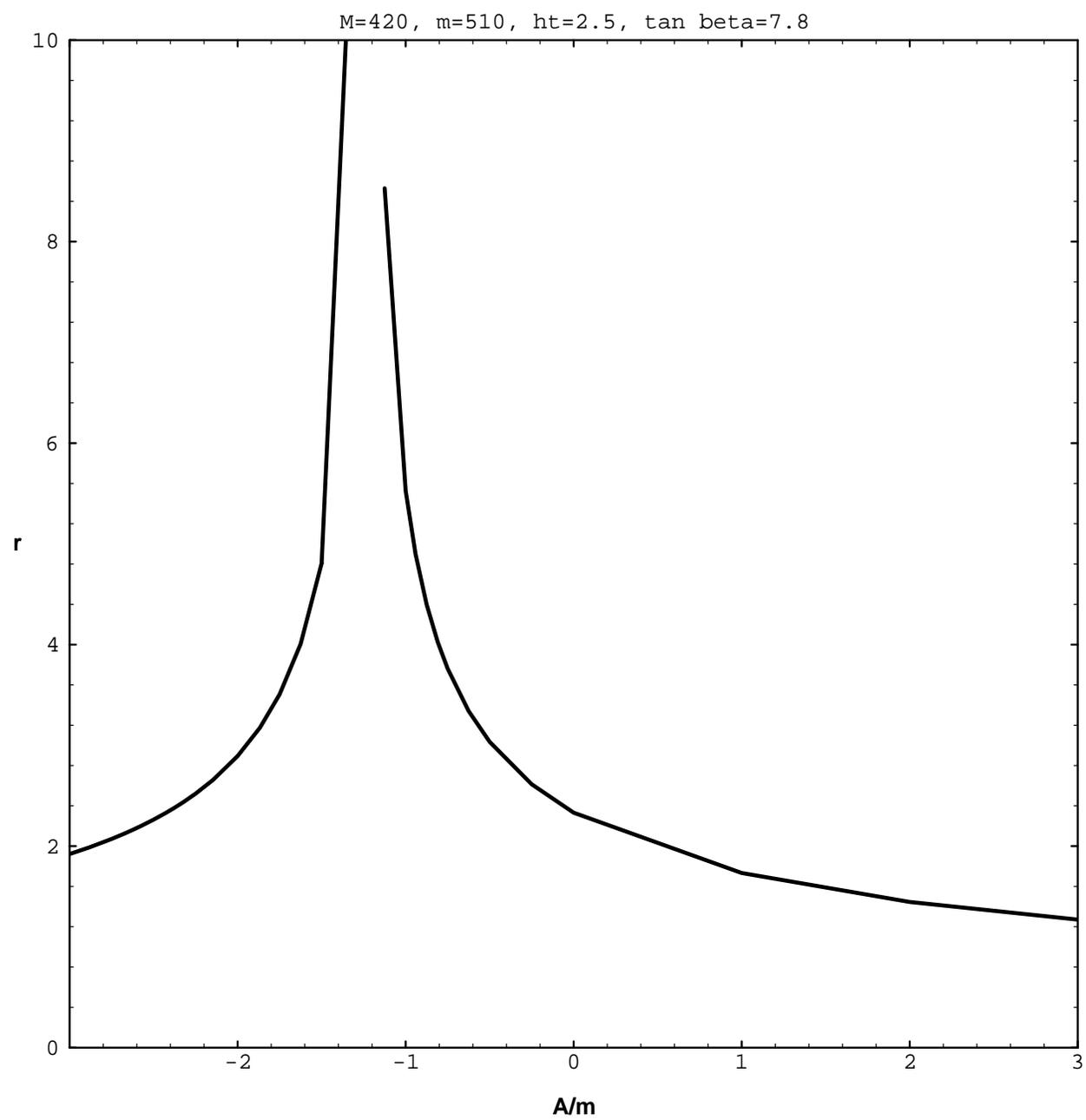
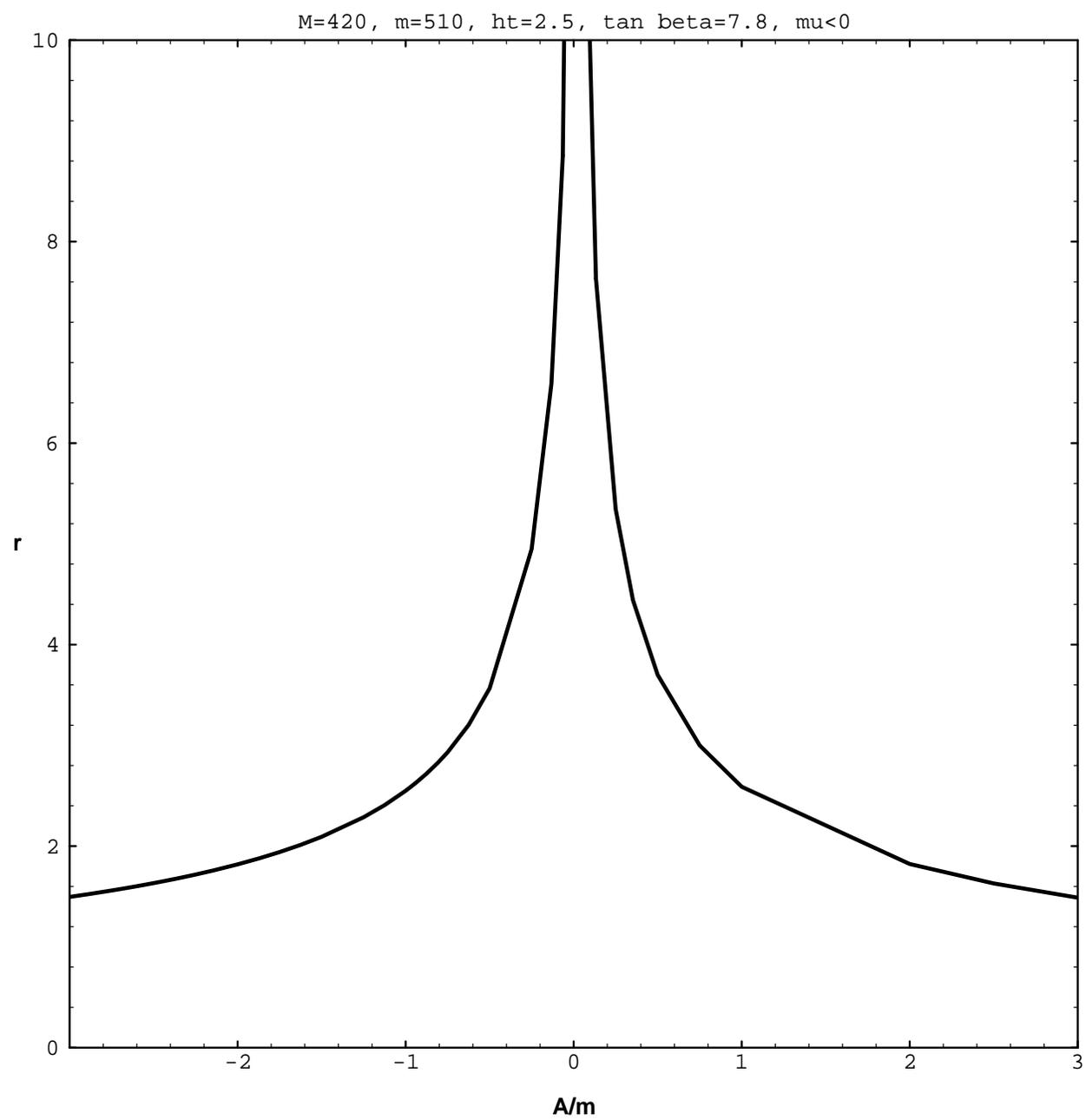
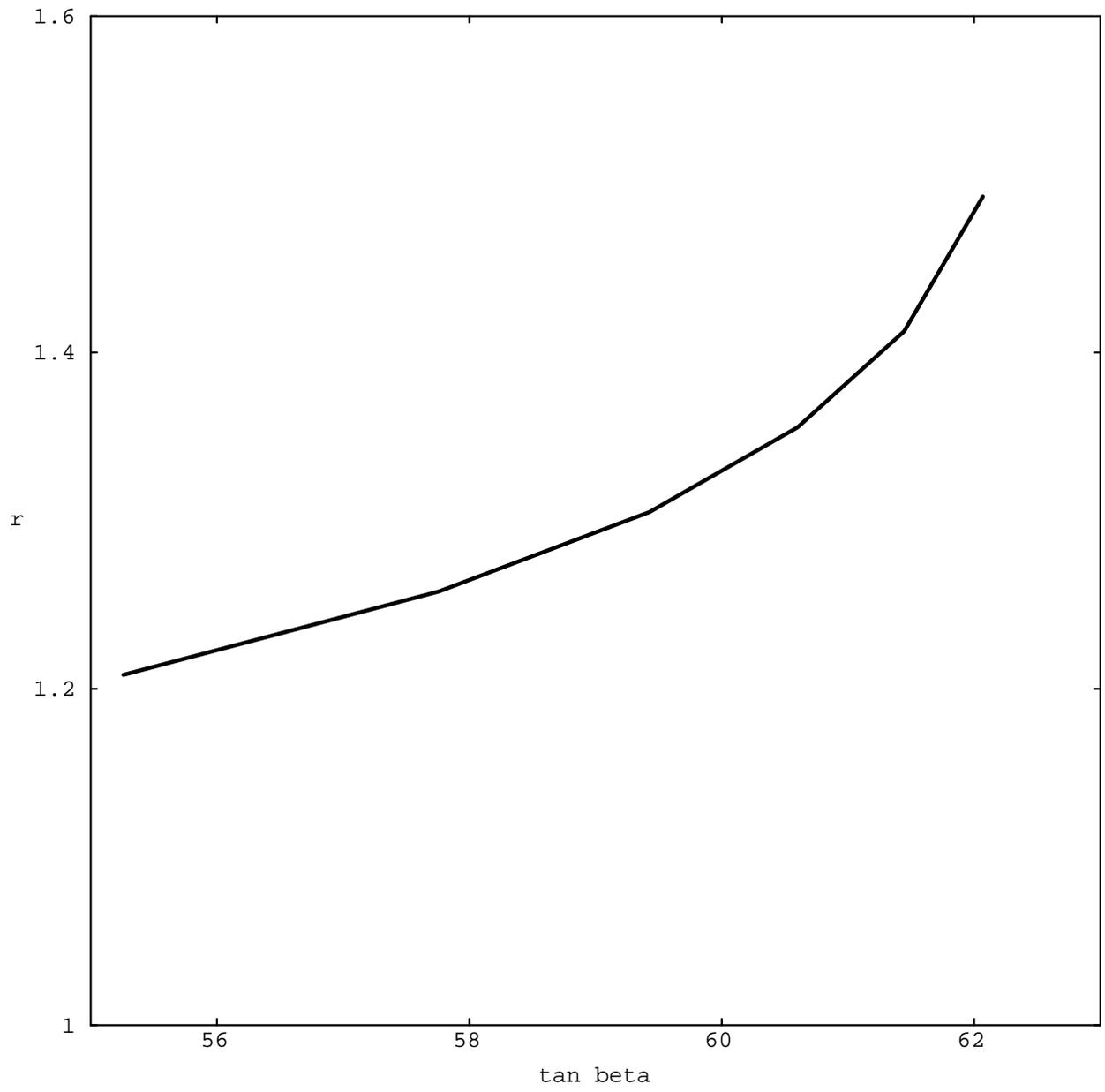
\end{document}